\documentclass[aps,reprint,amsmath,amssymb,superscriptaddress,floatfix]{revtex4-1}

\usepackage[colorlinks,citecolor=blue,linkcolor=blue]{hyperref}
\usepackage{graphicx}
\usepackage{dcolumn}
\usepackage{bm}
\usepackage{color}
\usepackage{amsfonts}
\usepackage{txfonts}

\newcommand{\CNN}{Centre de Nanosciences et de Nanotechnologies, CNRS, Univ. Paris-Sud, Universit\'e Paris-Saclay, 91405 Orsay cedex, France}
\newcommand{\Exeter}{School of Physics, University of Exeter, Stocker Road, Exeter EX4 4QL, United Kingdom}
\newcommand{\INRIM}{Istituto Nazionale di Ricerca Metrologica, Strada delle Cacce 91, 10135, Turin, Italy}

\begin{document}
\title{Nonreciprocal flexural dynamics of Dzyaloshinskii domain walls}
\author{R. Soucaille}
\email{remy.soucaille@gmail.com}
\affiliation{\CNN}
\affiliation{\Exeter}
\author{F. Garcia-Sanchez}
\affiliation{\CNN}
\affiliation{\INRIM}
\author{J.-V. Kim}
\author{T. Devolder}
\author{J.-P. Adam}
\email{jean-paul.adam@u-psud.fr}
\affiliation{\CNN}

\date{\today}

\begin{abstract}
We revisit the description of ferromagnetic domain wall dynamics through an extended one-dimensional model by allowing flexural distortions of the wall during its motion. This is taken into account by allowing the domain wall center and internal angle to be functions of position in the direction parallel to the wall. In the limit of small applied fields, this model accounts for the nonreciprocity in the propagation of wall modes and their stability in the presence of the Dzyaloshinskii-Moriya interaction and in-plane magnetic field.
\end{abstract}

\maketitle

\section{Introduction}
The dynamics of ferromagnet domain walls encompasses a wide range of nonlinear phenomena. A prime example is Walker breakdown, which represents the transition between the steady state motion of a wall under applied fields or currents and the precessional regime in which the wall motion is oscillatory. Despite the complexity of the magnetic interactions and dynamics governing individual moments in a given material system, much of the salient features of domain wall motion can be captured by a one-dimensional model in which the wall center, $u(t)$, and internal wall angle, $\phi(t)$, are the only variables that describe the dynamics~\cite{Slonczewski:1973, Schryer:1974}. For example, the magnetic field dependence of the wall velocity predicted by the model has been observed in in-plane magnetized wires~\cite{beach_dynamics_2005}.

Despite the utility of the one-dimensional model, a large number of experimental observations cannot be accounted for by this description. For example, magnetic disorder can prevent the Walker transition to be attained or even identified clearly~\cite{metaxas_creep_2007}, and leads to a low-field creep regime where thermally-activated processes are dominant~\cite{lemerle_domain_1998}. Beyond the creep regime, other deviations from the one-dimensional picture have been observed in perpendicularly-magnetized ultrathin ferromagnets. Instabilities in the wall structure lead to plateaus in the velocity versus field curves~\cite{yamada_influence_2011, burrowes_low_2013}, which are largely driven by incoherent magnetization precession at the wall center~\cite{voto_effects_2016}. Due to this incoherent precession, N\'eel or Bloch lines are created within and move along these domain walls, leading to periodic annihilation events resulting in spin wave bursts~\cite{yoshimura_soliton-like_2015}. A number of previous works has addressed such shortcomings in different ways. For the dynamics of vortex walls in in-plane magnetized systems, an extended model has been developed that also account for the internal dynamics of the vortex in addition to the usual wall variables~\cite{Tretiakov:2008, Clarke:2008}. This approach is based on the method of collective coordinates, which provides a framework to incorporate internal degrees of freedom (such as spin waves) into the dynamics of the underlying spin texture~\cite{bouzidi_motion_1990, Helman:1991, le_maho_spin-wave_2009}. Other work have sought to account for flexural modes of the domain wall, which can be excited during propagation and can lead to clear deviations from the one-dimensional behavior~\cite{Gourdon:2013}. Evidence of wall flexing has been obtained in low moment ferromagnets~\cite{Balk:2011}.

In ultrathin ferromagnets, the proximity of a strong spin-orbit coupling material can give rise to an additional chiral interaction of the Dzyaloshinskii-Moriya form~\cite{fert_role_1980, fert_magnetic_1990, Crepieux:1998}. Besides favoring N\'eel-type domain walls at equilibrium~\cite{heide_dzyaloshinskii-moriya_2008, thiaville_dynamics_2012, chen_tailoring_2013, tetienne_nature_2015, benitez_magnetic_2015, gross_direct_2016}, this interfacial Dzyaloshinskii-Moriya interaction (iDMI) results in the asymmetric nucleation~\cite{pizzini_chirality-induced_2014} and growth of magnetic domains in perpendicularly-magnetized ferromagnetic films under in-plane applied fields, where domain wall propagation is affected in the creep, steady state, and precessional regimes~\cite{kabanov_-plane_2010, mihai_miron_fast_2011, thiaville_dynamics_2012, je_asymmetric_2013, emori_current-driven_2013, Ryu:2013, hrabec_measuring_2014, torrejon_interface_2014, lavrijsen_asymmetric_2015, soucaille_probing_2016, Jue:2016, lau_energetic_2016, ha_pham_very_2016}. With respect to internal modes, it has also been shown that spin wave channeling by domain walls can acquire a nonreciprocal character~\cite{garcia-sanchez_nonreciprocal_2014, garcia-sanchez_narrow_2015}, which is similar to behavior seen for magnetostatic spin wave modes induced by dipolar effects~\cite{damon_magnetostatic_1961}. Some theoretical work has been undertaken to explore how the iDMI affects the wall motion in the creep regime~\cite{Pellegren_Dispersive_Stiffness_2017}, but there remain open questions on its role of the dynamics of flexural modes.


Here, we introduce an intermediate model between a full micromagnetic description of a domain wall and the 1D model~\cite{thiaville_dynamics_2012}. We explicitly allow for a nonuniform propagation of the domain wall, where the spatial dependence of the wall center is taken into account. This model allows us to have a more complete description of the domain wall dynamics. We will focus on the bending motion of a straight domain wall with iDMI and in-plane field. To describe the dynamics of the domain wall, especially at long wavelength we model the evolution of the domain wall dynamics in system with perpendicular magnetic anisotropy and in the presence of iDMI. We also examine the effect of a small in-plane magnetic field and pinning potentials.

The article is organized as follows. In Section II, we compute the dynamics of the domain wall via a Lagrangian description of the magnetic texture. Our model is a direct extension of the 1D model where the magnetization is supposed rigid. In Section III, we compare the obtained result for the flexural motion with micromagnetic simulations. In Section IV, we discuss in more details the non-reciprocity of the flexural modes. Some concluding remarks are provided in Section V.

\section{Domain wall energy and dynamics}

\begin{figure}
\centering\includegraphics[width=8cm]{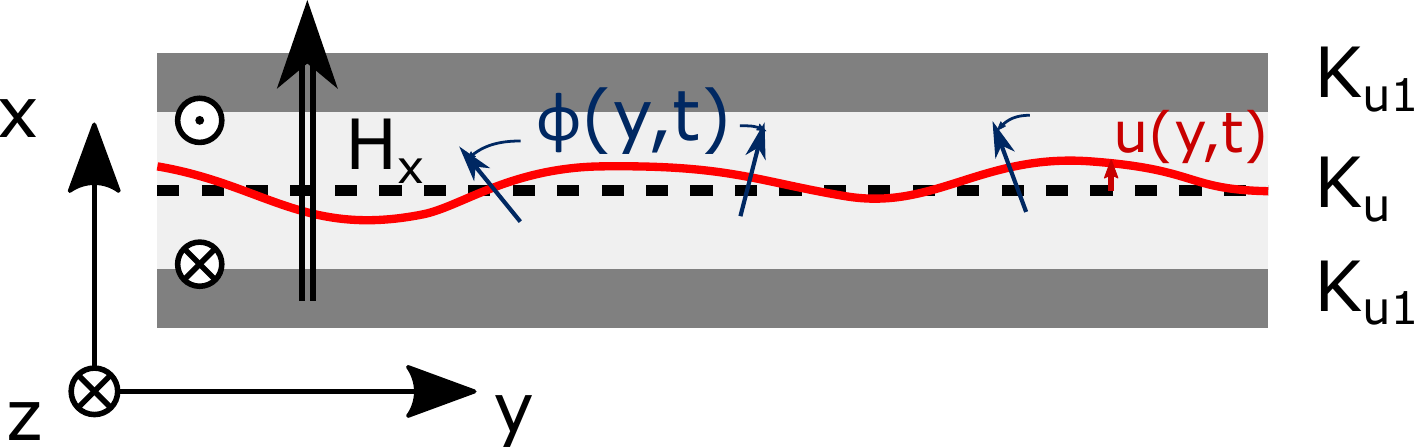}
\caption{\label{Fig_geometrie}Geometry for the domain wall dynamics. The domain wall is set at the center (dotted line). An anisotropy well where $K_{\rm u} < K_{\rm u,1}$ is used to model the effect of the pinning . The two function $\phi \left( y,t \right)$ and $u \left( y,t \right)$ parametrize respectively the in-plane angle and the out-of-plane component of the magnetization. The red line is a schematic of the domain wall position during an excitation.}
\end{figure}

We describe the energy and dynamics of a magnetic domain wall in an ultrathin film in this section. Within the micromagnetic description, the magnetization unit vector $\mathbf{m} = \mathbf{M}/M_s$ can be described by the two spherical angles $\theta$ and $\phi$,
\begin{equation}
\mathbf{m} \left(\theta , \phi \right) = \left(\sin\theta \cos\phi  ,  \sin\theta  \sin\phi  , \cos  \theta  \right).
\end{equation}
We consider a domain wall running along the $y$ direction that separates two magnetic domains along the $x$ direction. The magnetization within each domain is taken to point along the $z$ direction, perpendicular to the film plane. A schematic illustration of this geometry is given in Fig.~\ref{Fig_geometrie}. 
We assume that the component of the magnetization perpendicular to the plane, parametrized by the angle polar angle $\theta$, can be described by the usual profile,
\begin{equation}
\theta(x,y,t) =2 \tan^{-1} \left[ \exp \left( \frac{x-u\left(y,t\right)}{\Delta}\right) \right],
\label{equ_bloch_profile}
\end{equation}
where $u \left( y, t \right)$ represents the domain wall center displacement and $\Delta = \sqrt{A/K_\mathrm{eff}}$ is the domain wall width, which remains constant. $A$ is the exchange constant and $K_\mathrm{eff}$ is the perpendicular anisotropy constant. Both $\theta$ and $\phi$ are assumed to be uniform across the film thickness along $z$. We also assume that $\phi$ is uniform along the wall in the $x$ direction, $\phi = \phi \left(y, t\right)$ and that the domain wall remains close to a straight configuration ($\vert\vert \partial u / \partial y\vert\vert  \ll 1$).

We now describe the different contributions to the magnetic energy densities with this \emph{ansatz}. The effective perpendicular uniaxial anisotropy is given by
\begin{equation}
\mathcal{E}_{\rm anis} = K_{\rm eff}d \int \sin^2 \theta \, dx = \frac{1}{2} \sigma_0 d,
\label{equ_esurf_anis}
\end{equation}
where $d$ is the film thickness and $\sigma_0 = 4 \sqrt{A K_\mathrm{eff}}$ is the Bloch wall energy. This energy is independent of the displacement $u$ and the angle $\phi$. The exchange interaction energy density is given by
\begin{align}
\mathcal{E}_{\rm ex} &  = A d \int\left[ \left(\nabla \theta \right)^2 + \sin^2\theta \left( \nabla \phi\right)^2 \right] dx, \nonumber  \\
					&=  \frac{1}{2}\sigma_0 d \left[ 1 + {\left( \dfrac{\partial u}{\partial y} \right)}^2 + \Delta^2 \left( \dfrac{\partial \phi}{\partial y} \right)^2 \right]. \label{equ_esurf_ex}
\end{align}
The (positive) squared derivative terms are a reminder that the exchange energy favors a uniform configuration along the transverse (i.e., $y$) direction. The sum of these two terms give the usual domain wall energy, $\sigma_0$, with additional contributions to the wall elastic energy, proportional to $( \partial_y u )^2$ and $(\partial_y \phi)^2$, which arise from small deformations from the straight wall profile.

In perpendicularly magnetized films described by these two magnetic energies only, domain walls are of the Bloch type which minimize volume dipolar charges. Deviations from this profile can appear when other interactions are present. First, proximity of a strong spin-orbit coupling material to the ferromagnetic film induces an iDMI, which can be described by~\cite{bogdanov_chiral_2001, thiaville_dynamics_2012}
\begin{align}
	\mathcal{E}_{\rm D}   &  = D d \int \left( m_z \left( \nabla \cdot \mathbf{m} \right) - \left( \mathbf{m} \cdot \nabla\right)  m_z \right)  dx, \nonumber\\
					&= - \pi D d \, \left( \cos \phi - \frac{\partial u}{\partial y} \sin \phi \right).\label{equ_esurf_DMI}
\end{align}
One can rewrite this equation as a function of $\phi+ \tan^{-1} \left(\partial u /\partial y \right)$ and the length of the domain wall which shows that the iDMI effective field~\cite{thiaville_domain-wall_2006} is always normal to the domain wall. Second, applied magnetic fields can also modify the wall profile. The effect of a magnetic field $\mathbf{H}$ can be separated into two parts, an in-plane ($//$) and an out-of-plane ($\perp$) component. The out-of-plane component results in domain wall displacement and contributes to the total energy density through the Zeeman interaction as,
\begin{equation}
	\mathcal{E}_{Z,\perp} = - 2 \mu_0 M_{\rm s} H_z u d, \label{equ_esurf_HOOP}
\end{equation}
In contrast, the in-plane component leads to changes in the internal structure $\phi$ of the domain wall. In the limit of a small applied in-plane magnetic field, $\sqrt{H_x^2 + H_y^2} \ll H_{K}$, where $H_K = 2 K_\mathrm{eff}/\mu_0 M_s$ is the effective anisotropy field, the associated Zeeman term is
\begin{equation}
	\mathcal{E}_{Z,//} = - \pi \mu_0 M_{s} \Delta d \left(H_{\rm x} \cos \phi - H_{\rm y} \sin \phi\right). \label{equ_esurf_HIP}
\end{equation}
Note that as compared to the one dimensional model~\cite{thiaville_domain-wall_2006} we have an additional term proportional to $\partial_y u$ in the iDMI energy, which is absent in the Zeeman energy. While part of the iDMI energy can still be assimilated to an effective magnetic field along the $x$ direction, we note that the $y$ component of the effective field is linked to gradients in the domain wall displacement $u$, $\partial_y u$. This result may provide a theoretical basis for asymmetric domain growth that has been observed experimentally in the presence of in-plane fields \cite{kabanov_-plane_2010,je_asymmetric_2013,hrabec_measuring_2014,lavrijsen_asymmetric_2015,Pellegren_Dispersive_Stiffness_2017,Balk_Simultaneous_Control_2017}.

We approximate the dipolar interaction with a transverse anisotropy term, $\mathcal{E}_{\perp}$. We assume that the domain wall profile varies slowly compared to the domain wall width and consider flexural modes in the long wavelength limit, $k \Delta \ll 1$. Under this approximation, we can write
\begin{equation}
	\mathcal{E}_\perp = 2 K_\perp \Delta d \cos \left[ \phi + \tan^{-1} \left( \frac{\partial u}{\partial y}\right)\right]^2,	\label{equ_esurf_perp}
\end{equation}
where $K_\perp = \left(\ln 2\right) \mu_0 M_{\rm s}^2 d/ \pi \Delta$  accounts for the difference in dipolar energy between the N\'eel and Bloch wall profiles~\cite{kim_intrinsic_2016}. We note that the dipolar interaction, like the iDMI, leads to a coupling between the internal angle $\phi$ and deformations in the wall, $\partial_y u$. The dipolar interaction favors the Bloch profile, where the magnetization at the wall center is tangent to the domain wall, while the interfacial iDMI favors the N\'eel profile, where the magnetization at the wall center is normal to the domain wall. Finally, domain wall pinning due to material inhomogeneities need to be accounted for realistic systems. This can be introduced by assuming a quadratic potential well of the form,
\begin{equation}
	\mathcal{E}_{\rm pin} \approx \frac{1}{2} \kappa u^2 d, \label{equ_esurf_pin}
\end{equation}
where $\kappa$ characterizes the strength of the pinning potential.

For systems with uniform properties along the $y$ axis, such as the line anisotropy defect shown in Fig.~\ref{Fig_geometrie}, we can assume that the equilibrium domain wall profile is also uniform along this direction. As such, $\phi(y) = \phi_0$ and all spatial derivatives in the domain wall position are vanishing, $\partial u /\partial y = 0$. In this case, the equilibrium domain wall angle $\phi_0$ can be found by minimizing the total energy. The equilibrium angle $\phi_0$ can be determined by minimizing the total energy. In the case where the in-plane field is applied along the $x$ direction, $\phi_0$ is given by~\cite{kim_intrinsic_2016}
\begin{equation}
\phi_0 = 
\begin{cases}
\pi \qquad \; \mu_0  \pi M_{\rm s} H_x <  \pi D/\Delta - 2 K_\perp   \\
0 \qquad \; \mu_0  \pi  M_{\rm s} H_x >   \pi D/\Delta + 2 K_\perp   \\
2 \tan^{-1} \left[ \sqrt{\frac{2K_\perp + \pi D/\Delta + \mu_0  \pi M_{\rm s} H_x}{2K_\perp - \pi D/\Delta - \mu_0  \pi M_{\rm s} H_x}}\right] \; \textrm{otherwise}
\end{cases}
\end{equation}
Depending on the in-plane field the domain wall structure transforms from a right handed N\'eel wall to a left handed N\'eel passing through a mixed N\'eel/Bloch wall.

We now discuss the dynamics of the domain wall as an elastic line using Lagrangian formalism to derive the equations of motion of the domain wall. For spin dynamics, the Lagrangian density for the spherical angles $\theta$ and $\phi$ can be written as ~\cite{kim_chapter_2012, Braun:2012},
\begin{equation}
\mathcal{L}  = \frac{M_s d}{\gamma} \int \left(  \dot{\phi}\left(1 -  \cos\theta \right)  - \mathcal{E} \left[\theta, \phi \right]  \right) dx,
\label{equ_lagrangien}
\end{equation}
where the first term on the right hand side is the Berry phase term and $\mathcal{E} \left( \theta , \phi \right)$ is the total energy density of the domain wall, which is the sum of the different contributions given in Eqs.~(\ref{equ_esurf_anis}) to (\ref{equ_esurf_perp}). By using the domain wall ansatz in Eq.~(\ref{equ_bloch_profile}), integrating over $x$ and removing terms which are not relevant for the dynamics of the system, we find 
\begin{equation}
\mathcal{L}  = - 2 \frac{M_{\rm s} d}{\gamma} \dot{\phi} u - \mathcal{E} \left[u, \phi \right],
\end{equation}
Gilbert damping can be accounted for through the dissipation function
\begin{equation}
\mathcal{W}_{\rm G} = \alpha \frac{M_s d}{2 \gamma} \int \left(\dot{\theta}^2 +  \dot{\phi}^2 \sin^2\theta \right) dx,
\end{equation}
which, with the same ansatz, leads to the density
\begin{equation}
\mathcal{W}_{\rm G} =  \frac{\alpha M_{s} \Delta d}{\gamma} \left( \frac{\dot{u}^2}{\Delta^2}  + \dot{\phi}^2\right).
\end{equation}

The equations of motion correspond to the usual Euler-Lagrange equations,
\begin{align}
\frac{d}{dt}\frac{\partial \mathcal{L}}{\partial (\partial_t u)} + \frac{d}{dy}\frac{\partial \mathcal{L}}{\partial (\partial_y u)} - \frac{\partial \mathcal{L}}{\partial u} + \frac{\partial \mathcal{W}_{\rm G}}{\partial (\partial_t u)}
&= 0, \\
\frac{d}{dt}\frac{\partial \mathcal{L}}{\partial (\partial_t \phi)} + \frac{d}{dy}\frac{\partial \mathcal{L}}{\partial (\partial_y \phi)} - \frac{\partial \mathcal{L}}{\partial \phi} + \frac{\partial \mathcal{W}_{\rm G}}{\partial (\partial_t \phi)}
&= 0.
\label{equ_lagrangian_formel}
\end{align}
Explicitly, this leads to the set of coupled nonlinear differential equations,
\begin{widetext}
\begin{multline}
        \frac{2 M_{\rm s}}{\gamma} \frac{\partial \phi}{\partial t} +\frac{M_{\rm s} \alpha}{\gamma \Delta} \frac{\partial u}{\partial t} - \sigma_0 \frac{\partial^2 u}{\partial y^2} - \pi D \frac{\partial \phi}{\partial y} \cos \phi - 2 \mu_0 M_s \pi H_z  + \kappa u - \\
         K_\perp \left( \frac{\partial \chi}{\partial y} \sin{2\left(\chi +\phi \right)} \sin 2 \chi - 2\cos{2\left(\chi+\phi\right)} \left(\frac{\partial\phi}{\partial y}+\frac{\partial\chi}{\partial y}\right) \cos^2 \chi \right) = 0,   
        \label{equ_complet1} 
\end{multline}
\begin{equation}
        -\frac{2 M_{s} }{\gamma} \frac{\partial u}{\partial t} +\frac{M_{\rm s} \alpha}{\gamma }\Delta \frac{\partial \phi}{\partial t}- 4 A \Delta \frac{\partial^2 \phi}{\partial y^2} + \pi D \left( \frac{\partial u}{\partial y} \cos \phi + \sin \phi \right) - K_\perp \Delta \sin \left( 2\chi +2\phi \right) + \mu_0 M_s \pi \Delta \left( H_x \sin \phi - H_y \cos \phi \right) = 0,    
         \label{equ_complet2}
\end{equation}
\end{widetext}
where $\chi \equiv \tan^{-1} (\partial u/\partial y)$. Note that we recover the usual one-dimensional domain wall model~\cite{thiaville_dynamics_2012} when the spatial derivatives in $u$ and $\phi$ are set to zero.

\section{Dispersion relation of the flexural mode}

In this section we focus on the the dispersion relation for the flexural dynamics of the domain wall derived from the Lagrangian formalism in the previous section. We limit our study to the case where the slope of the wall is small, $\partial u / \partial y \ll 1 $, for small wavelengths, $ k \Delta \ll 1$, and for small applied in-plane fields along the normal to the domain wall, $H_{//} = H_x \ll H_{\rm K_{\rm eff}}$. A first order series expansion on the set of two coupled differential equations is done over $u \left(y,t \right)$ and $\phi \left( y,t \right)$ around $\phi = \phi_0$ and $u = 0$ where $\phi_0$ is the equilibrium angle of the magnetization. We then look for the propagating solutions of theses equations to get the dispersion relation $\omega \left( k\right)$, which is found to be
\begin{equation}
\omega \left( k \right) = \Omega_{\rm NR}+ \sqrt{\Omega_u \Omega_\phi},
\label{equ_disperlation}
\end{equation}
where $\Omega_{\rm NR}$ is linear with respect to the wave vector and is responsible for non-reciprocal propagation,
\begin{equation}
\Omega_{\rm NR} = \omega_{{\rm D},k} \cos \phi_0 -k \Delta \omega_\perp \cos 2 \phi_0. \label{equ_omega_nr}
\end{equation}
$\Omega_u$ describes the stiffness of the domain wall position, 
\begin{equation}
\Omega_u = \omega_k+\omega_{\rm pin}-\left(k \Delta\right)^2 \omega_\perp \cos 2 \phi_0,
\label{equ_omega_u}
\end{equation}
and $\Omega_\phi$ describes the stiffness of the angle $\phi$ of the magnetization, 
\begin{equation}
\Omega_\phi = \omega_k+ \frac{\omega_{{\rm D},k}}{k \Delta} \cos \phi_0 + \omega_{H,x} \cos \phi_0 - \omega_\perp \cos 2 \phi_0.
\label{equ_omega_phi}
\end{equation}
The angular frequencies in Eqs.~(\ref{equ_omega_nr}) to (\ref{equ_omega_phi}) are given by~\cite{garcia-sanchez_narrow_2015}
\begin{align}
	\mathcal{\omega}_k              &=  2\frac{\gamma A}{ M_{\rm s}} k^2, \\
	\mathcal{\omega}_{\mathrm{D,}k} &=  \frac{\pi}{2}\frac{\gamma D}{M_{\rm s}} k, \\
	\mathcal{\omega}_\perp          &=  \frac{\gamma K_\perp}{ M_{\rm s}}, \\
	\mathcal{\omega}_{H,x} &=  \frac{\pi}{2} \gamma \mu_0 H_x, \\
	\mathcal{\omega}_{\rm pin}      &=  \frac{1}{2}\frac{\gamma \Delta}{M_{\rm s}}\kappa. \label{equ_omega_pinning}
\end{align}
We note that this description provides a more accurate treatment of the dipolar interaction at long wavelengths in comparison to previous work~\cite{winter_bloch_1961, braun_fluctuations_1994, garcia-sanchez_narrow_2015}, since we take into account fluctuations in the wall position through the term $\partial_y u$.

\begin{figure}
\centering\includegraphics[width=8cm]{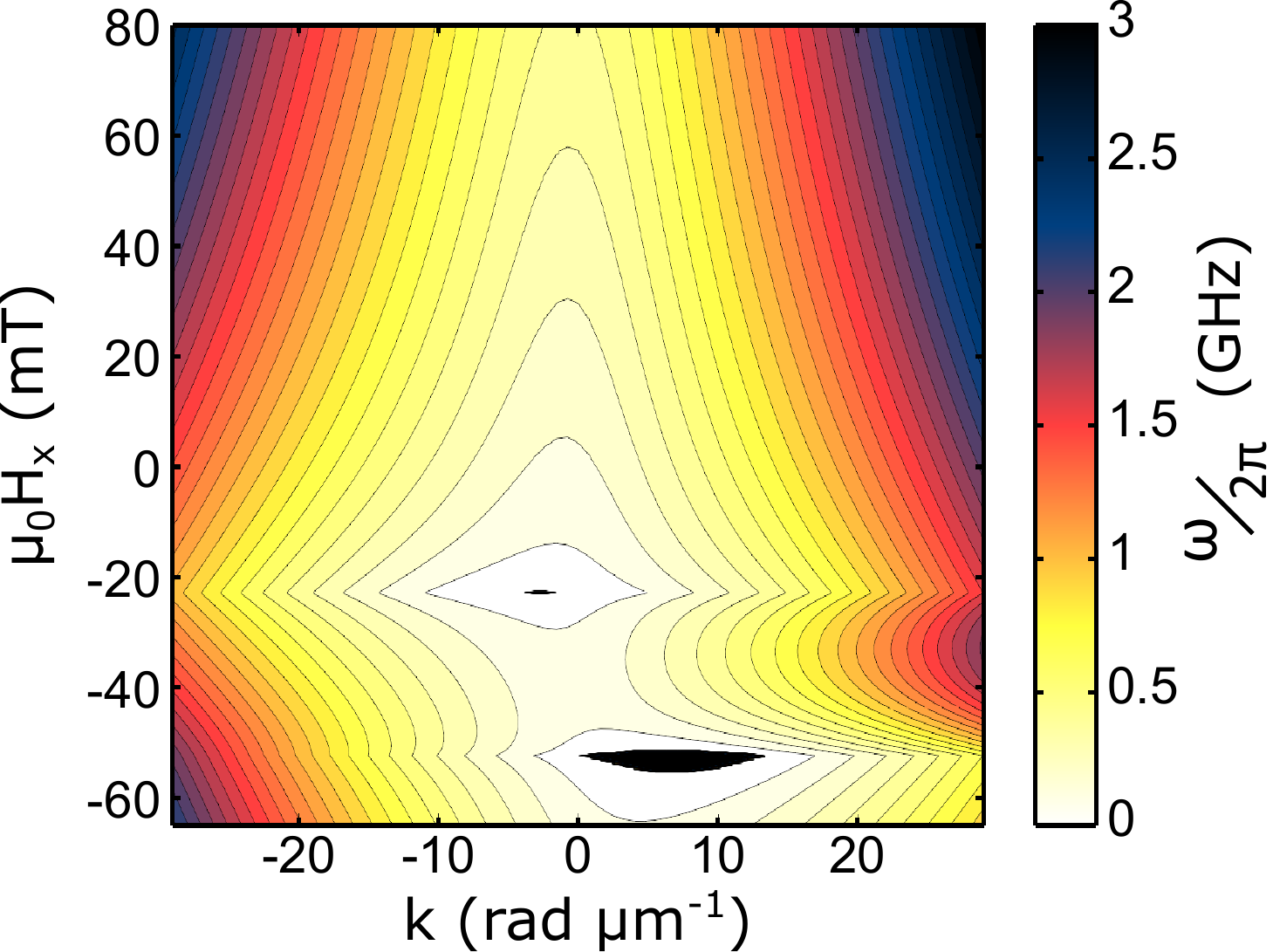}
\caption{\label{Fig_champ_veconde_frequ_model}Dispersion relation of the flexural motion of the domain wall with respect to the in-plane field $\mu_0 H_{x}$ normal to the domain wall and the wave vector $k_y$ which propagates along the domain wall [from Eq.~(\ref{equ_disperlation})]. The wall is pinned in an anisotropy well with a width of 80~nm. Contour lines are separated by 100~MHz and the black areas close to $-20$~mT and $-50$~mT represent the ($\mu_0 H_{x}$, $k_y$) space range in which the frequencies are negative, indicating straight domain wall instabilities.}
\end{figure}

The variation of the flexural mode frequency with applied in-plane field, $H_x$, and the wave vector, $k$, is presented in Fig.~\ref{Fig_champ_veconde_frequ_model}. 
The dispersion relation allows us to identify instabilities in the straight domain wall configuration assumed as the equilibrium profile. Instabilities occur when the mode frequency vanish, which typically corresponds to a change in the equilibrium state. We note that the uniform mode, $k=0$, which corresponds to the uniform displacement of the domain wall, is not necessarily the eigenmode with the lowest energy. This can be seen in Fig.~(\ref{Fig_champ_veconde_frequ_model}) where there are two field intervals over which the straight domain wall is unstable. Similarly to the case where the iDMI is strong enough to create maze domain patterns~\cite{garcia-sanchez_narrow_2015}, this instability occurs for one propagation direction of the flexural model.

Unlike this case, here the domain wall energy is still positive and the sign of wave vector which leads to negative frequency depends on the chirality of the domain wall and on $D$. This instability is shown in Figure~\ref{Fig_champ_veconde_frequ_model}, where the frequency reaches negative values for wave vectors between 0 and $10~\mathrm{\mu m^{-1}}$ at around -50~mT and for $-5\,\mathrm{\mu m}^{-1} < k < -1\,\mathrm{\mu m}^{-1}$ at $-20$~mT. For an applied in-plane field value close to the field value corresponding to the iDMI strength a straight wall is stable. This result is surprising since one expects faceting of the wall when the in-plane field compensates the iDMI field~\cite{lau_energetic_2016,Pellegren_Dispersive_Stiffness_2017}. We do not find such an instability here. This is linked to the initial state that we have considered, where $\phi_0$ is constant in the initial configuration whereas for a curved domain wall $\phi_0$ can vary along the domain wall. The two side of facets has opposite slope ($\partial u / \partial y$) and opposite angle $\phi$. To create facets the angle $\phi$ needs to go continuously from positive to negative value. For this, the domain wall needs to overcome an energy barrier which is linked to the dipolar interaction: a straight domain wall configuration is a metastable state. The dipolar energy has two opposite effects, it broadens the region where a straight wall is metastable and it tends to increase the stability of the domain wall in this region by increasing the energy barrier. For example, if we do not consider the dipolar interaction (Eq.~\ref{equ_esurf_perp}) the two regions of instability in Fig.~\ref{Fig_champ_veconde_frequ_model} merge into one region which is centered close to the iDMI effective field.

We compared the analytical model with full micromagnetic simulations using \textsc{Mumax3}~\cite{vansteenkiste_design_2014}. The simulation consists of a domain wall pinned in an anisotropy well where the uniaxial anisotropy is increased by 10\% outside the well. The width of the anisotropy well is taken to be 80~nm. The dynamics are induced by a pulsed magnetic field, with a time dependence given by a sinc function, along the $z$ direction localized at the center of the frame. The simulation geometry is given by a rectangular window with dimensions of $8 \; \rm \mu m$  by $125$~nm, which is discretized with 2048 $\times$ 64 finite difference cells. The layer thickness is 1~nm and periodic boundary conditions are applied along the $y$ direction (along the domain wall). The saturation magnetization is taken to be $M_s=788$~kA/m, the exchange stiffness $A = 22.5$~pJ/m, the uniaxial anisotropy $K_0 = 641.5$ kJ/m$^3$, the iDMI $D = 0.28$ mJ/m$^2$ and the Gilbert damping constant $\alpha = 0.015$. These values represent the material system W(3 nm)/CoFeB(1 nm)/MgO, as discussed in Ref.~\onlinecite{soucaille_probing_2016}.

\begin{figure}
\centering\includegraphics[width=8cm]{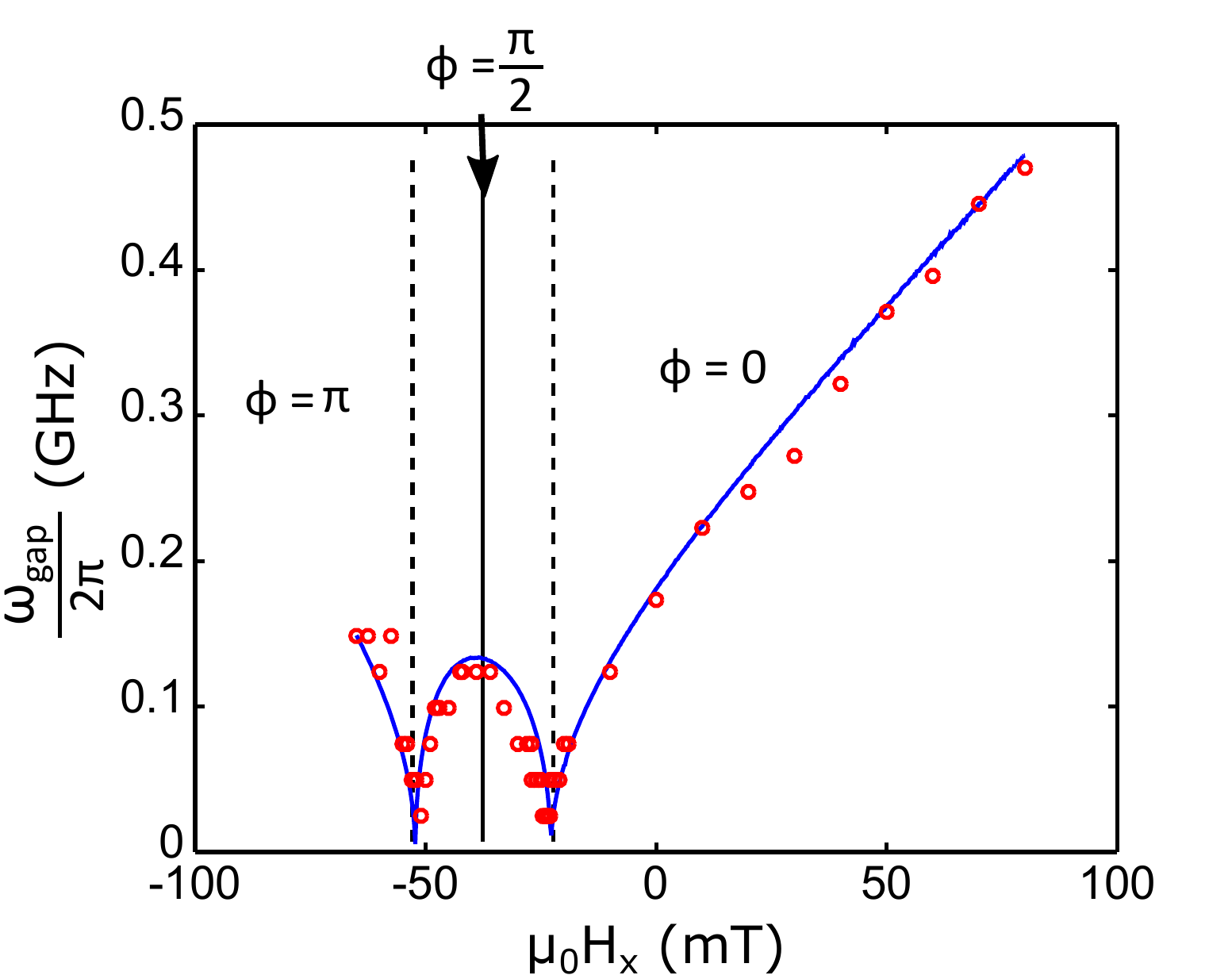}
\caption{Influence of the in-plane field on the frequency gap at $k=0$. The solid line is derived from the model and circles are result of micromagnetic simulations. The vertical lines show the different internal angles $\phi$ of the domain wall at equilibrium. \label{Fig_Evolution_fres_vs_IPfield}}
%
%
\end{figure}

A comparison is given in Fig.~\ref{Fig_Evolution_fres_vs_IPfield} for the finite frequency gap for the uniform $k=0$ mode, which appears due to the uniform pinning potential as illustrated in Fig.~\ref{Fig_geometrie}. 
An expression for this gap is given by
\begin{equation}
	\omega_{\rm gap} = \sqrt{\omega_{\rm pin}\left(\frac{\omega_{{\rm D},k}}{k \Delta} \cos \phi_0 - \omega_\perp \cos 2 \phi_0\right)}.
\end{equation}
Besides setting the pinning potential to zero (\textit{i.e.} $K_u = K_{u1}$), the frequency gap can also vanish when the applied in-plane field leads to changes in the domain wall profile. This can be seen in Fig.~\ref{Fig_Evolution_fres_vs_IPfield} at around $\mu_0 H_x = -20$ mT and $-50$ mT, where transitions toward the two different chiral N\'eel domain wall states occur. The figure shows that good agreement between the analytical model and the micromagnetics simulations is obtained, both for the critical fields and the gap frequencies.

\begin{figure}
\centering\includegraphics[width=8cm]{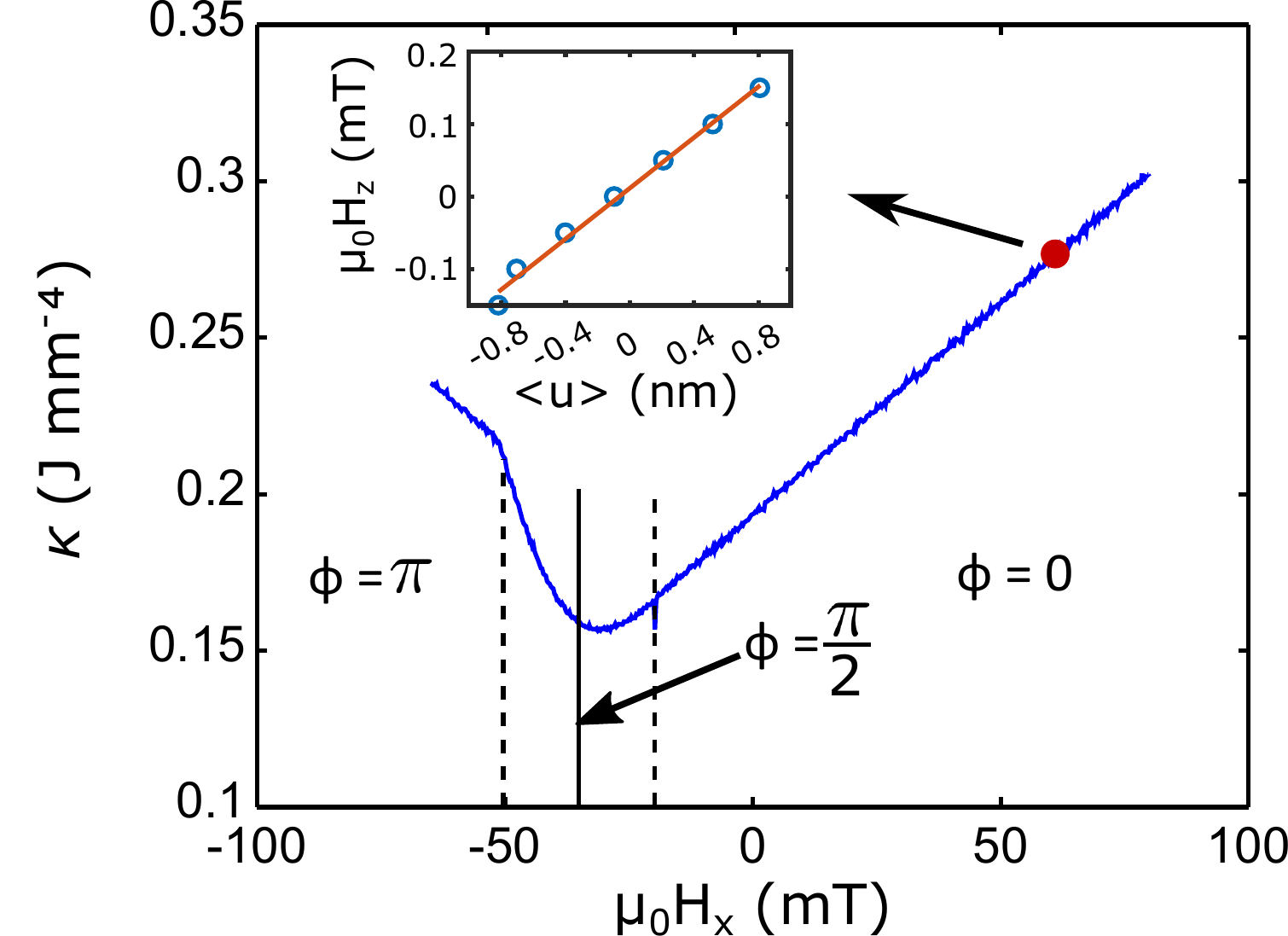}
\caption{Variation of the pinning constant $\kappa$ as a function of the in-plane applied field. The vertical lines delimit the different domain wall configurations for the angle $\phi$.  The inset shows one example of the fitting between the out-of-plane field with respect to the domain wall position. We deduce the position pinning from this linear fit and with the equation~\ref{equ_pinvsu}.  \label{Fig_Evolution_pinning}}
\end{figure}

In the micromagnetic simulations pinning also occurs due to the presence of edges close to the domain wall. In systems with iDMI the boundary conditions are modified~\cite{rohart_skyrmion_2013,garcia-sanchez_nonreciprocal_2014} which can contribute to pinning. In order to estimate the anisotropy well contribution to the pinning potential, we analyze the wall displacement with an out-of-plane field. We focus on the quadratic pinning potential, Eq.~(\ref{equ_esurf_pin}), and the Zeeman interaction, Eq. ~(\ref{equ_esurf_HOOP}). We minimize these two energy terms. The pinning potential is then related to the out-of-plane field and the wall position by,
\begin{equation}
\kappa = \frac{2 \mu_0 M_{\rm s} H_z}{u}.
\label{equ_pinvsu}
\end{equation}
The pinning potential is then obtained by a linear fit to the positive field branch, as shown in the inset of Figure~\ref{Fig_Evolution_pinning}.
The pinning exhibits a variation that is similar to the domain wall width~\cite{kim_intrinsic_2016}. When the applied field is opposed to the wall chirality the wall width is slightly reduced~\cite{kim_intrinsic_2016} in our simulation, as the width of the anisotropy well is larger than the wall width the pinning is reduced. From the pinning strength the resonant frequency can be estimated.

\begin{figure}
\centering\includegraphics[width=7.5cm]{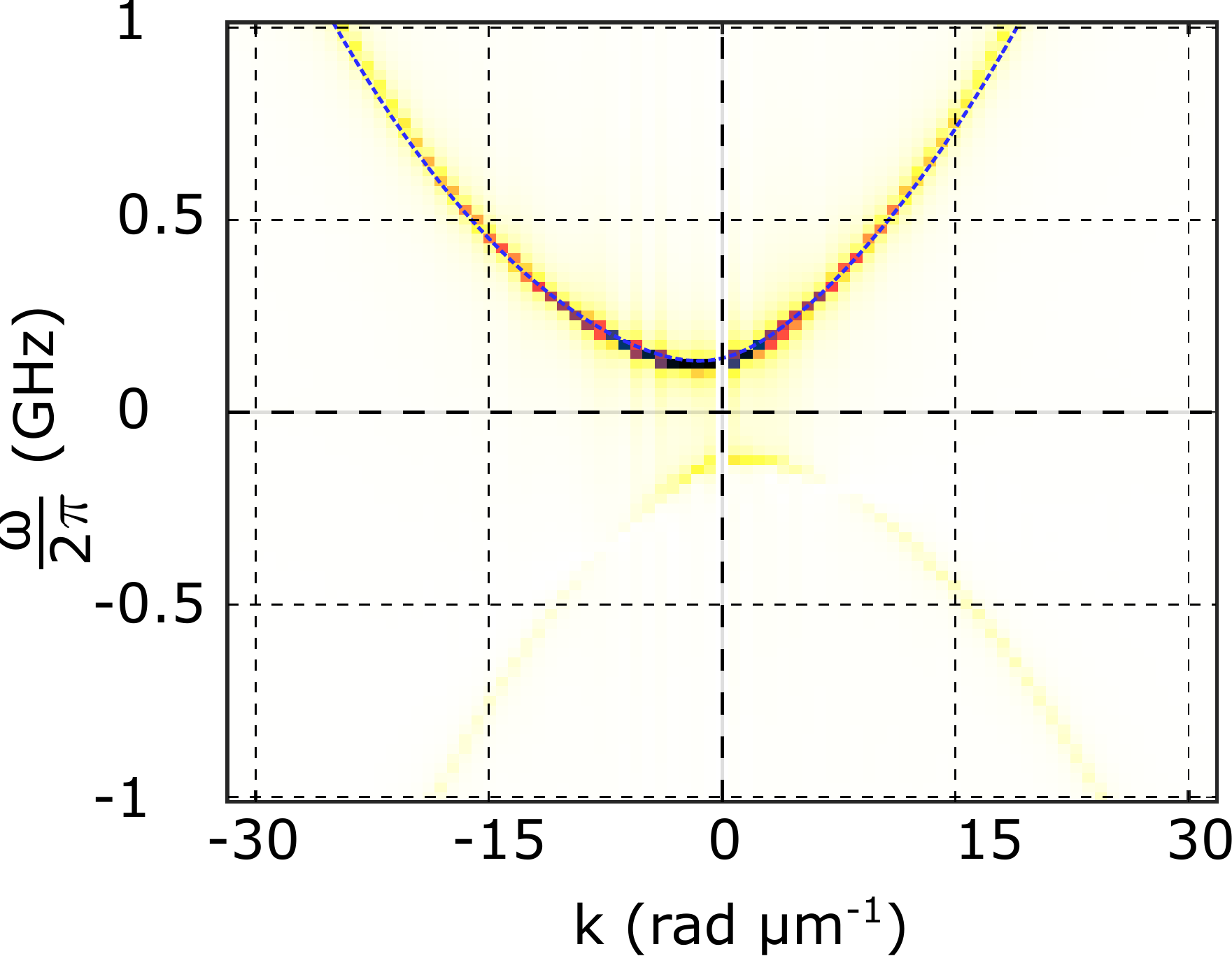}
\caption{\label{Fig_sus_-30mT_freq_k}Detail of the dispersion relation for an applied field $\mu_0 H_x = -10 \; \textrm{mT}$. It corresponds to the 2D Fourier transformation of the function $u+ i \Delta \phi$ with respect to the time and the "$y$" variable. The blue dotted line corresponds to the dispersion relation given by Eq.~(\ref{equ_disperlation}).}
\end{figure}

In order to compute the full dispersion relation we extract the domain wall position, $u\left(y,t\right)$, and the angle of the magnetization at the domain wall center, $\phi \left( y,t \right)$. Linear extrapolation is used to obtain values of the displacement smaller than the cell size. The dispersion relation is then computed from a two-dimensional Fourier transform of the complex-valued function $u + i \Delta \phi$, as shown in Fig.~\ref{Fig_sus_-30mT_freq_k}. The remaining branch in the frequency wave vector space leads to negative frequency when instabilities occur. At this applied field the energy difference between a straight domain wall and the relaxed state is small. 
In this section we have seen that an in-plane field allows to modify the resonant frequency. To properly describe the dynamic of flexural motion of the domain wall in domain walls at relatively low frequencies, i.e. in the sub-GHz range, we have to consider the effect of pinning in the flexural modes.

\section{Nonreciprocity of the flexural mode}
\begin{figure}
\centering\includegraphics[width=7.5cm]{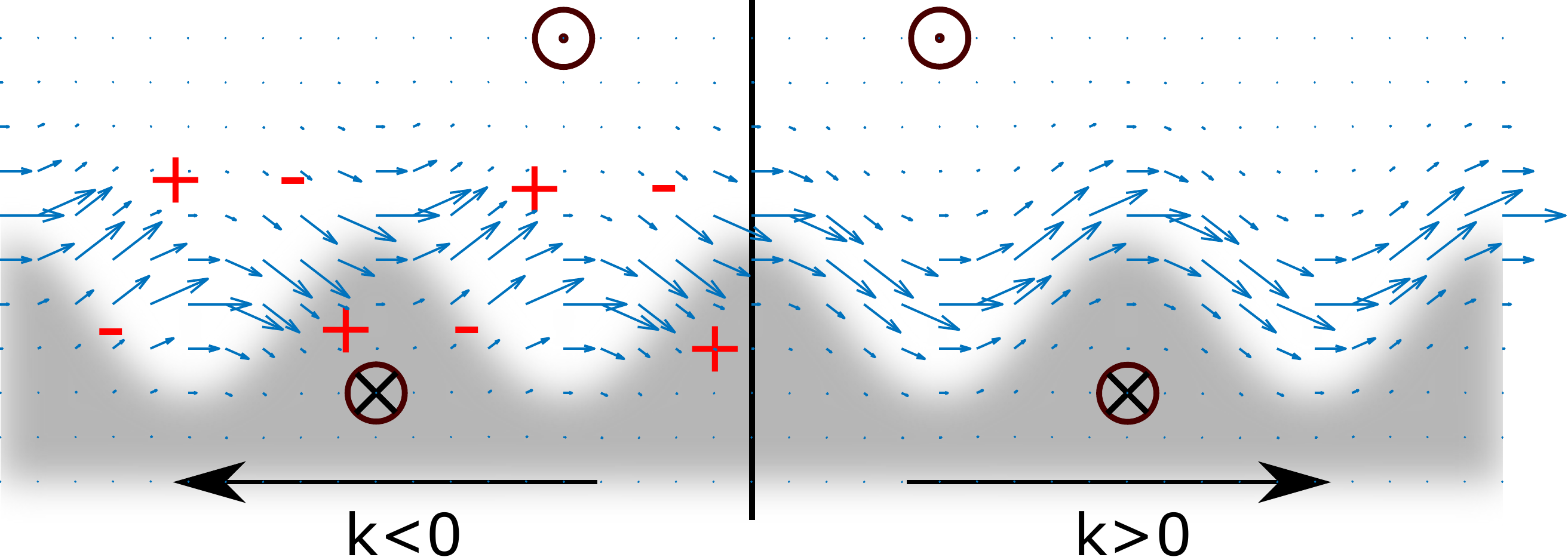}
\caption{Illustration of the magnetization state for two opposite propagation directions. The ``+'' and ``-'' represent the excess of virtual magnetic charges which appear for $k<0$. \label{Fig_non_reciprocity_illustration}}
\end{figure}

In this section we discuss the nonreciprocal behavior of the flexural mode. In the absence of dipolar interactions~\cite{garcia-sanchez_narrow_2015}, the frequency difference between two counterpropagating spin waves is proportional to the iDMI constant $d$. However, as seen in Eq.~(\ref{equ_omega_nr}), dipolar interactions also lead to an additional nonreciprocity. While the contribution from the iDMI is periodic with respect to $\phi$, the nonreciprocity related to the dipolar interaction is $\pi$-periodic [see Eq.~(\ref{equ_disperlation})] and vanishes when $\phi = \pi / 4$. This is related to the fact that the dipolar contribution does not depend on the domain wall chirality but rather on the presence of volume dipolar charges (i.e., whether the wall is of the Bloch or N\'eel type), whereas the iDMI is sensitive to the domain wall chirality. If we consider a Bloch wall and only the in-plane component of the magnetization, the two counterpropagating waves lead to two different configurations as represented in Figure~\ref{Fig_non_reciprocity_illustration}.

\begin{figure}
\centering\includegraphics[width=8cm]{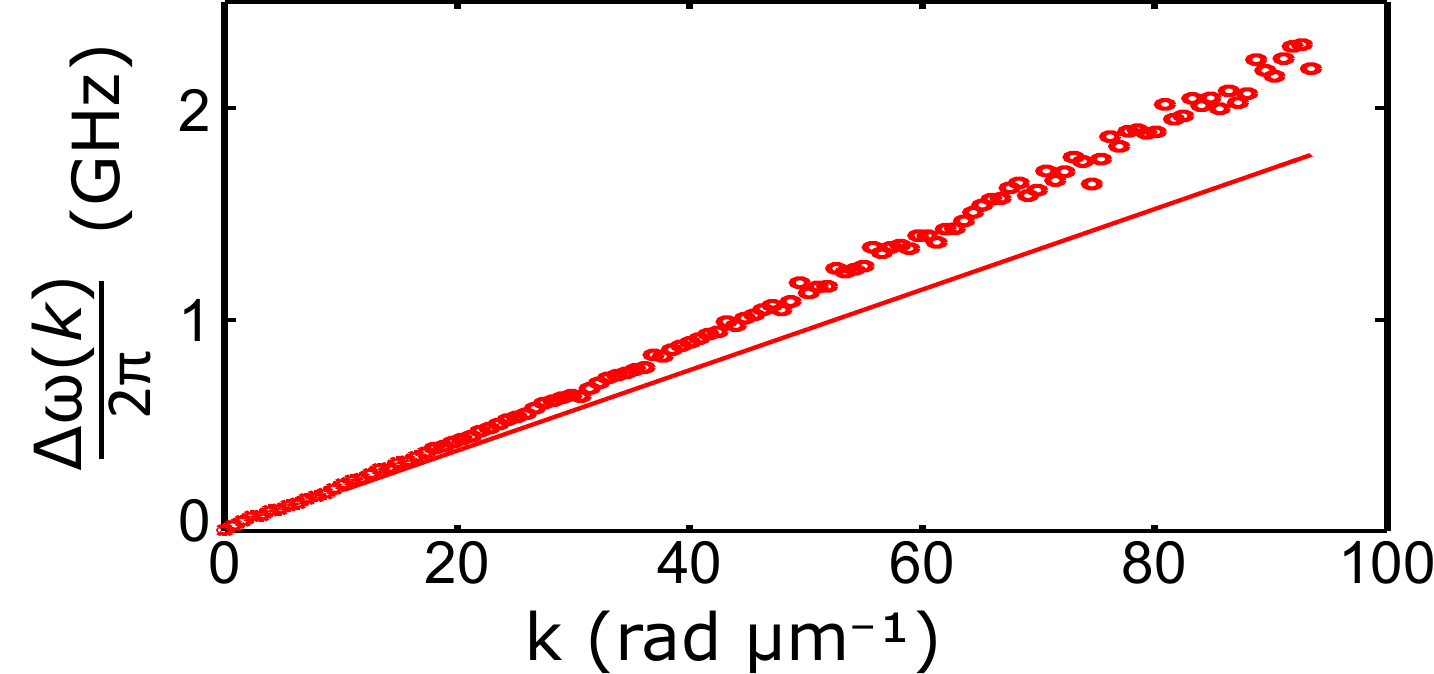}
\caption{Evolution of the non-reciprocity with respect to the wave vector with applied in-plane field. Circle are results from micromagnetic simulation, the red line corresponds to the approximation where the demagnetizing field is taken locally.
\label{Fig_non_reciprocity}}
\end{figure}

In one case the in-plane magnetization is tangent to the domain wall position, $k > 0$,  while this is not the case for the opposite direction, $k < 0$. This leads to a difference in the dipolar interaction for this two cases; when the magnetization is tangent to the domain wall the dipolar energy is reduced. To describe accurately this nonreciprocity we cannot neglect the dipolar interaction if the iDMI is small. In this case the dipolar energy and the iDMI have a similar magnitude,
\begin{equation}
\pi^2 D \sim 2 \ln 2 \mu_0 M_{\rm s}^2 d.
%
%
\end{equation}
For $k \Delta \ll 1$, the function  $\Delta \omega =\omega (k) - \omega(-k)$ is linear with respect to the wave vector due to both the dipolar and the iDMI contributions. As seen in Figure~\ref{Fig_non_reciprocity} the non-reciprocity does not follow the same linear trend for larger wave vectors.
For $k \Delta \gg 1$, the dipolar contribution to the nonreciprocity saturates and the slope of the function $\Delta \omega$ is dominated by the contribution from the iDMI.

\begin{figure}
\centering\includegraphics[width=8cm]{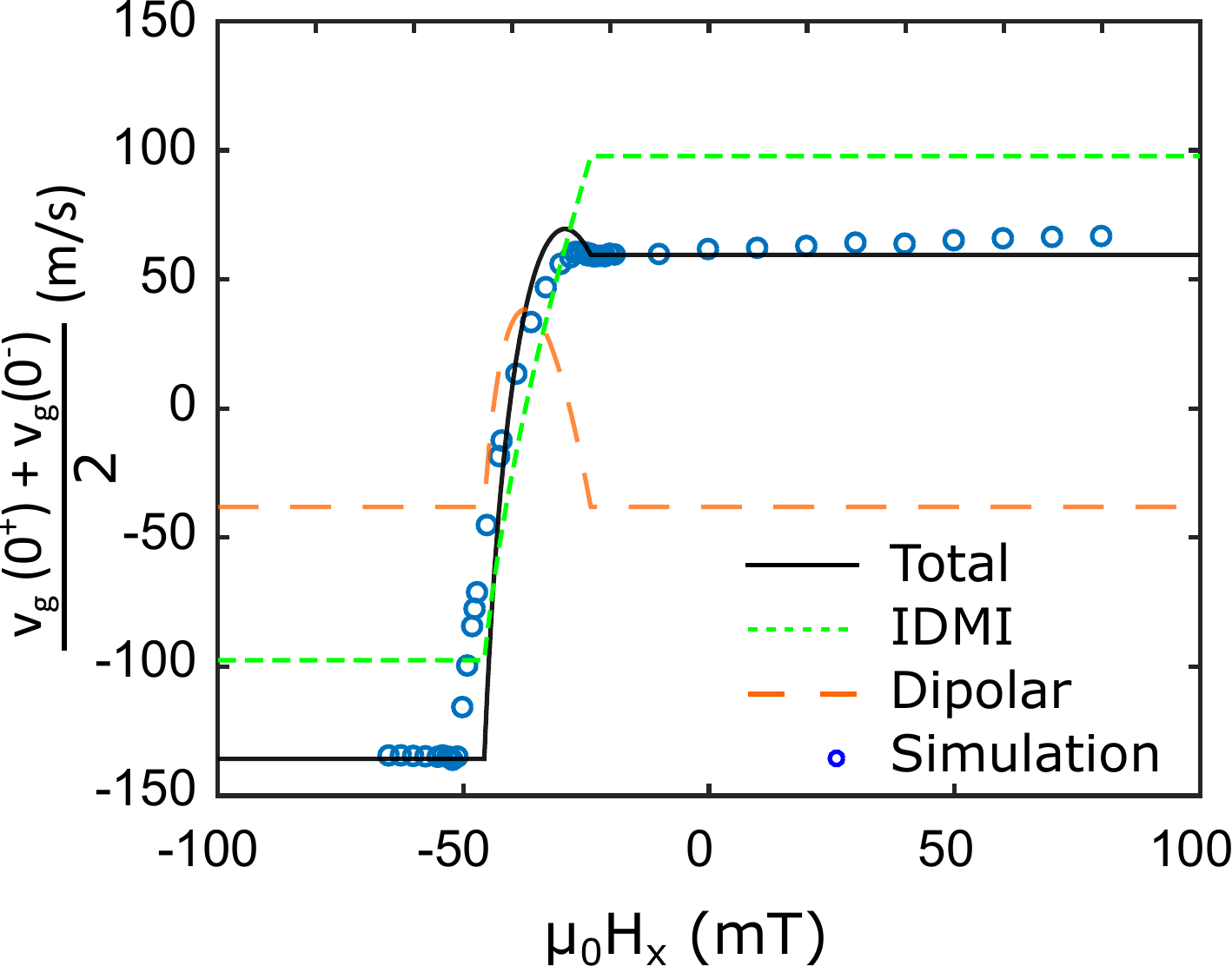}
\caption{Evolution of the group velocity for small wavevector with respect to in-plane field $\mu_0 H_x$ along the $x$ axis. Circle are results from micromagnetic simulation, the black line correspond to the model. The two dotted lines correspond to the iDMI and to the dipolar field contribution to the wall excitation non-reciprocity. \label{Fig_non_reciprocite_W3}}
\end{figure}

In order to have a closer look to the contributions of the different energy terms to the nonreciprocity we have considered the group velocity defined by $v_{\rm g} \left( k \right) = \partial \omega / \partial k$ as a function to the in-plane magnetic field. To match our assumption, we consider only small wave vector for the group velocity and we finally plot the mean group velocity for two opposite direction (\textit{i.e.} $ (v_{\rm g} \left( 0^+ \right) + v_{\rm g} \left( 0^- \right))/2$) in Figure~\ref{Fig_non_reciprocite_W3} versus the applied in-plane field.
The two components of the nonreciprocity shown (dotted lines) depends on the in-plane field $H_x$, which provide a means to modify the nonreciprocity of the domain wall flexural modes. Micromagnetic simulations are in good agreement with the model in the long wavelength limit. As the non-reciprocity is mainly related to the domain wall structure, as soon as the domain wall saturates in one configuration a change in the applied in-plane field does not longer modify the frequency non-reciprocity. Our analytical model allows an accurate description of the non-reciprocity in this case. This long wavelength regime might be probed in Brillouin light scattering experiments~\cite{di_direct_2015} where the wavelength probed is in the order of few hundred nanometers.

\section{Conclusion}
We have developed an analytical model for domain wall dynamics beyond the one-dimensional model by allowing for inhomogeneous displacements of the domain wall under the influence of a magnetic field. In our model we have supposed that the domain wall structure remains rigid while the domain wall profile is allowed to bend. This description of a flexible domain wall allows for a quantitative analytical description of the flexural mode propagation in the domain wall at long wavelengths. We show that both the interfacial Dzyaloshinskii-Moriya and dipolar interactions contribute to the nonreciprocity of the flexural mode. We have studied the influence of an in-plane field on the domain wall dynamics, which allows the band gap induced by domain wall pinning to be tuned. This field can also induce instabilities in the flexural mode propagation, which take place at the Bloch-N\'eel transition. This change in the internal wall structure also modifies the direction and the amplitude of the nonreciprocity of the flexural modes.

\begin{acknowledgments}
The authors acknowledge fruitful discussions with Yves Henry and Matthieu Bailleul. This work was partially supported by Agence National de la Recherche (France) under contract No. ANR-16-CE24-0027 (SWANGATE).
\end{acknowledgments}

\bibliography{REFS_VF}

\end{document}